# Software Evolution Understanding: Automatic Extraction of Software Identifiers Map for Object-Oriented Software Systems

Ra'Fat AL-msie'deen, and Anas H. Blasi

*Abstract*—Software companies usually develop a set of product variants within the same family that share certain functions and differ in others. Variations across software variants occur to meet different customer requirements. Thus, software product variants evolve overtime to cope with new requirements. A software engineer who deals with this family may find it difficult to understand the evolution scenarios that have taken place over time. In addition, software identifier names are important resources to understand the evolution scenarios in this family. This paper introduces an automatic approach called Juana's approach to detect the evolution scenario across two product variants at the source code level and identifies the common and unique software identifier names across software variants source code. Juana's approach refers to common and unique identifier names as a software identifiers map and computes it by comparing software variants to each other. Juana considers all software identifier names such as package, class, attribute, and method. The novelty of this approach is that it exploits common and unique identifier names across the source code of software variants, to understand the evolution scenarios across software family in an efficient way. For validity, Juana was applied on ArgoUML and Mobile Media software variants. The results of this evaluation validate the relevance and the performance of the approach as all evolution scenarios were correctly detected via a software identifiers map.

## I. INTRODUCTION

SOFTWARE product variants often evolve from the initial version [1]. Each variant meets specific requirements defined by the customer. However, these software product variants usually share some common code and differ in other code [2]. When software product variants become numerous, comparing the code of the initial and the latest version is a solution to define the common and unique code for each variant in order to understand software evolution [3].

In fact, to understand the evolution of variant code, the software engineer asks important questions such as, why the code related to version A is deleted and why the code related to version B is added? Is it due to bug fixing, coping with changes or to add/remove functionality? Comprehension of software evolution scenarios requires an understanding of the existing software products. Existing evidence shows that successful coder uses software structure as well as software identifier names to discover software product [42]. With as much costs, effort, and time spent on understanding software evolution scenarios, there is a serious need for automated tools to help discover and comprehend today's huge and complex software variants evolution.

The main issue in software evolution analysis is the identification of specific changes that happen across numerous releases of a software product [35]. After the emergence of Lehman's laws of software evolution [43], it has been well comprehended that software system has to be modified to changing requirements and environments or it becomes increasingly less helpful. Software changes are generally known as an essential part of a software's life cycle [44]. Thus, recently numerous approaches have been developed to help software developers in understanding evolution scenarios in huge complex software products [1, 10, 13].

Software identifier names (e.g., packages, classes, attributes and, methods) are important software understanding sources [4, 5]. Identifier names across product variants need to be studied in order to understand the evolution scenarios in those variants. The main purpose of this paper is to help software engineers to compare the identifier names of two software product variants. This comparison aims to understand the evolution scenarios between these versions through source code changes. However, software engineer detects common and unique identifier names across software variants via software identifiers map. In fact, the main contribution of this research is to extract the identifiers map for two similar versions of the software product.

The identifiers map defines the names of the common identifiers for both versions, as well as the names of the unique identifiers for each software product.

*Juana's approach* identifies the common and unique software identifier names between two Object-Oriented (OO) software variants. The common software identifiers are present in two software variants. Furthermore, the unique software identifiers have presented in one software variant, while absent in another one. Juana computes common and unique software identifier names by comparing software variants to each other. However, the final result of Juana is the software identifiers map, which is a visual presentation of software variant identifier names, presented the common and unique identifier names between two product variants.

The novelty of Juana is that it exploits all software identifier names of product variants to identify the common and unique identifier names across those variants. Juana separates the identifiers of two product variants into two subsets, the common identifiers set, and the unique identifiers set. Indeed, common identifier names appear in all variants, while the unique identifier names appear in one variant but not all variants.

Manual reverse engineering of common and unique identifier names for software product variants is a tedious process, time-consuming, and needs large efforts. Supporting this process would be of great aid. This study suggests an automatic approach to extract evolution scenarios from two product variants. Juana is based on the identification of the implementation of this evolution scenario among identifier names of the source code. These identifier names form the initial search space. Juana uses Formal Concept Analysis (FCA) to reduce this search space. FCA divides the set of identifier names into two subsets, the common identifier names set, and the unique identifier names set. Then, it separates unique identifier names set into small subsets that each contain identifier names that are held uniquely by a certain software variant.

Juana is detailed in this paper as follows: Section II discusses all related work to Juana's contribution. Section III gives an overview of Juana. Section IV illustrates the software identifiers map extraction process in detail. Section V presents the experiments that were conducted to validate Juana's proposal. Finally, section VI presents a conclusion and provides a future work.

## II. RELATED WORK AND COMPARISON WITH JUANA

This section presents the related work to Juana's contributions. Also, it offers a brief overview of the different approaches and shows the need to propose Juana.

By going through software evolution literature review, it has been found that there is limited related work to the software evolution using software identifier names. In fact, some researchers were used FCA to study the variability across product variants, and others were compared the whole code of two products to extract unique feature implementations.

Al-Msie'deen *et al.* [12, 48] used FCA as part of their automatic feature model extraction technique. In their work, FCA was used to identify the common source code block and variable code blocks (*i.e.,* variability) across a collection of OO software product variants. In fact, Juana deals only with two OO software variants and identifies common and unique software identifier names (*i.e.,* identifiers map).

Rubin and Chechik [13] proposed in their paper an approach to locating distinguishing features of two software variants developed via code cloning. Their approach identified distinguishing features – those are presented in one software but not all software variants. Thus, the unique features are implemented in the unshared parts of the software code. Juana finds unique and common software identifier names across two software variants.

Fluri *et al.* [35] presented a change distilling tool called *CHANGEDISTILLER[1]*, a tree differencing procedure for fine-grained code change detection. CHANGEDISTILLER tool identifies fine-grained code changes among subsequent releases of software classes, based on calculating variances of their abstract syntax trees. As a result, software engineers obtained a set of elements that are new or changed in product P2, compared to product P1.

Source code variation has proven itself to be a continuing research issue essential to product variants analysis [36]. Raghavan *et al.* presented *Dex* [37], a tool for mining code variations among C source files. When software variants evolved over time, its UML models also are evolving. Kelter *et al.* identified differences between UML models [38]. Sager *et al.* [39] presented an approach to extract similarities across different software classes based on abstract syntax trees.

Kuhn [40] introduced a lexical approach to automatically recover labels from software components. His approach can be applied to compare software component terms with each other in order to understand components evolution. An approach was presented Anslow *et. al.* [41] to show the evolution of words in class names in Java release 1.1 and release 1.6. The authors showed the evolution history in a combined word cloud that holds terms from both versions of software systems. The cloud displays a comparison of the class names among Java version 1.1 (red color) and version 1.6 (blue color). Release 1.1 consists of 477 classes and release 1.6 consists of 3777 classes. A word cloud is an inspiring visualization method as it displays how the words used in software class names have changed among different releases of software variants. Word cloud shows that all of the words used in release 1.1 have also been used in release 1.6. There are a number of extra words used in release 1.6 which is to be predictable being a more recent release.

Al-Msie'deen and Blasi [1] proposed an automatic approach called (Iris) to study the software when it evolves over time, its code remains to grow, change and become extra complex. The novelty of their approach is the exploitation of the product variants to examine the influence of software evolution on the software metrics. Based on the mined software metrics, it has been found that the approach hypothesis is confirmed by the calculated metrics. Horwitz [45] presented an approach to compute semantic and textual differences between two software products. Baxter *et al.* [46] described a tool for code clone detection. However, the code clone tool relies on the abstract syntax tree.

---

[1] https://www.ifi.uzh.ch/en/seal/research/tools/changeDistiller.html

Several studies [10, 14] were used the FCA technique to study the variability across software family. However, FCA used as a clustering technique to extract common parts and unshared parts of the variant's source code, but FCA is not already used to provide a clear, simple, and accurate visual presentation of the software identifier names for two software variants as in Juana's approach.

### III. APPROACH OVERVIEW

This section presents the main concepts and hypotheses used in Juana's approach for extracting the software identifiers map from software variants source code. In addition, this section gives an overview of the software identifiers map extraction process. It also describes the toy example that illustrates the remaining of the paper.

The main goal of this research is to understand software evolution across two software variants. The Successful software variants may have been presented many years ago with a new version released every year. Furthermore, the software product is changed to reflect changing customer requirements over time. For large and long-lifetime software systems that are developed by a software company for customers, systems must evolve to meet changing customer requirements [15]. However, it is important to understand software evolution.

Juana is concerned with re-documenting software variants to make them easier to comprehend and change. The variants are documented through the map of identifiers extracted by Juana's approach. Juana extracts the software identifiers map of two OO software product variants. So, the software identifiers map shows the common and unique identifiers across product variants. By browsing and exploring the identifiers map, the programmer can see the changes in the code during the evolution of the software. In addition, changes in the software identifiers are clearly visible on the extracted map.

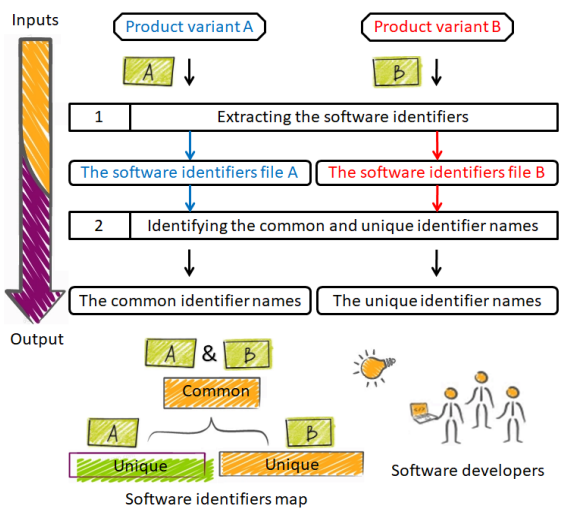

Fig. 1. The software identifiers map extraction process

[2] https://sites.google.com/site/ralmsideen/tools

The software identifiers map extraction process takes the variants' source code as input. The first step of this process aims to identify software identifiers based on the static code analysis. Second, identifies the common and unique software identifiers across two product variants based on FCA. Figure 1 shows the software identifiers map extraction process. Juana relies on a software identifiers map to determine the common and unique identifier names.

As an illustrative example, this paper considers two variants of the drawing shapes software family[2] [1, 16]. The first version of the drawing shapes software allows software engineers to draw three different kinds of shapes (*i.e.,* line, oval, and rectangle). The second version allows engineers to draw three different kinds of shapes (*i.e.,* line, round rectangle, and 3D rectangle). In fact, this toy example is used to better explain some parts of this paper. Juana only uses the source code of software variants as input but does not know the common and unique software identifier names in advance.

Figure 2 shows the common and unique identifiers between two product variants. Juana uses FCA as a clustering technique to find the common and unique identifiers across two product variants. The reason behind this choice is that the FCA technique expresses the wanted map artifact. The reader who is interested in FCA can find more information in many studies [6-10]. Based on two OO software product variants, Juana extracts all software identifiers based on the static code analysis [11] as a first step. Then, Juana uses the FCA to identify the common and unique identifier names (*i.e.,* software identifiers map) across software variants.

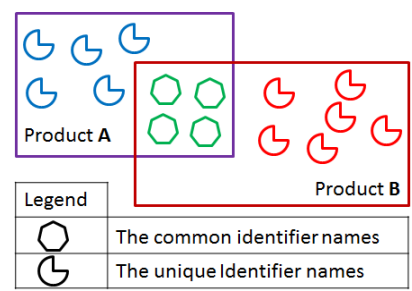

Fig. 2. The common and unique identifiers of two product variants

Juana identifies the common and unique identifier names across two OO software variants. However, Juana introduces the term of software identifiers map, which is an artifact gathering and viewing the common and unique identifier names across software variants. The main objective of Juana's approach is to help the software engineers understand the evolution that has occurred across product variants at the source code level. In addition, Juana's approach is the only current approach that studies evolution scenarios between two software products by exploiting software identifier names.

### IV. THE SOFTWARE IDENTIFIERS MAP EXTRACTION PROCESS

This section describes the software identifiers map extraction process in detail. However, the suggested approach extracts software identifiers map in two steps as detailed in the following section.

*A. Extracting the Software Identifiers*

The first step of the software identifiers map extraction process aims to extract all software identifier names for product variants. Juana static code parser extracts all software identifier names from software variants source code. As inputs for this step, Juana accepts only two software variants source code. The outputs of this step are two code files. However, for each software variant, there is a code file contains all software identifier names (*i.e.,* package, class, attribute, and method). The extracted code stored as XML files and the extracted file contains main OO identifiers in addition to main code relations such as inheritance, method invocation, and attribute access.

*B. Identifying the Common and Unique Identifier Names*

The second step of the software identifiers map extraction process is the identification of the common and unique identifiers. The technique used to identify them depend on FCA [17 - 20]. Initially, a formal context, where objects are software product variants and attributes are software identifier names, is extracted. The corresponding AOC-poset is then generated. Table I shows the formal context for the drawing shapes software variants.

TABLE I
THE FORMAL CONTEXT FOR THE DRAWING SHAPES SOFTWARE PRODUCT VARIANTS

| Drawing shapes releases | MyRoundRectangle | MyLine | DrawingShapes | PaintJPanel | MyShape | MyRectangle | My3DRectangle | MyOval |
|---|---|---|---|---|---|---|---|---|
| Release 1 |  | × | × | × | × | × |  | × |
| Release 2 | × | × | × | × | × |  | × |  |

Figure 3 shows the AOC-poset for the formal context of Table I and represents the software identifiers map. In the formal context, the product family appears as row labels, while software identifiers appear as column labels. Furthermore, the cross sign indicates that the corresponding product contains this identifier name. The AOC-poset in Figure 3 shows three concepts. Each concept in the AOC-poset consists of two parts: the concept intent and the concept extent. However, the intent of each concept represents software identifier names common to two variants or unique for one product. For example, the intent of the top concept (*i.e.,* concept_2) contains software identifiers that are common to two variants. The intents of all remaining concepts (*i.e.,* concept_0 and concept_1) are unique software identifier names. For example, the intent of concept_1 is the unique identifiers for the second release of drawing shapes software. On other hand, the extent of each of these concepts is the product that has these identifiers in its code. For instance, the extent of concept_0 is the first release of drawing shapes software.

Based on the identifiers map (*i.e.,* the AOC-poset), the software engineer can browse the map from top to bottom to see the common identifier names of the two programs as well as their unique identifier names. This map helps software developers understand the evolution of the program. The upper concept contains common identifiers that have not changed during the evolution of the program. While the rest of the concepts show the changes that have occurred to the program's identifiers during its evolution. Juana's approach extracts five types of maps, the extracted maps cover all software identifiers (*i.e.,* packages, classes, attributes, and methods). In addition, Juana extracts a map containing all software identifiers (*i.e.,* identifiers map). Figure 4 shows packages, classes, and attributes map.

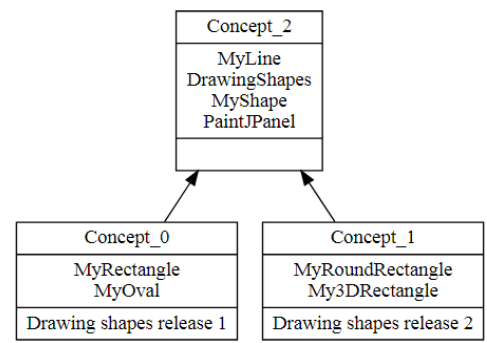

Fig. 3. The AOC-poset for the formal context of Table I

A quick look at the extracted maps shows that the packages and attributes of the software have not changed during its evolution, while there has been a change at the class level. In addition, some classes in the first version were deleted during the evolution of the program and other classes were added to the new version. However, these changes indicate that the program has evolved to meet the new requirements of the customer. The methods and identifiers map of drawing shapes variants are available on the Juana webpage [21].

The software identifiers map is very helpful for software developers to understand software evolution across two product variants at the source code level. Juana's approach can be used by software engineers when locating distinguishing identifiers – those are present in one variant but not all variants of the software family. Juana assumes that software variants are developed by the clone-and-own approach (*i.e.,* copy-paste-modify) [12].

## V. EXPERIMENT WORK

To validate the proposed approach, experiments ran on two real case studies: the mobile media [22] and ArgoUML [23]. Mobile media[3] software is a Java-based open-source application that manipulates media on mobile devices. ArgoUML[4] is a Java-based open-source software. ArgoUML tool includes support for all standard UML diagrams.

Table II summarizes the evolution scenarios in mobile media and ArgoUML software variants. The advantage of mobile media and ArgoUML variants is that they are well documented. Thus, the result of Juana's approach can be compared with the evolution scenarios documented in several studies [22, 23].

---

[3] http://www.ic.unicamp.br/~tizzei/mobilemedia/

[4] https://sdqweb.ipd.kit.edu/wiki/SPLevo/Case_Studies/ArgoUML-SPL

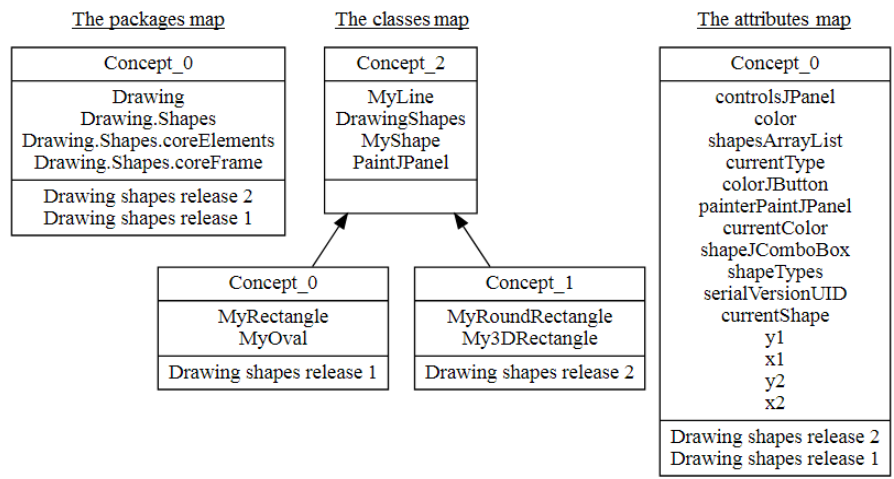

Fig. 4. The packages, classes, and attributes map extracted from drawing shapes variants

TABLE II
SUMMARY OF EVOLUTION SCENARIOS IN MOBILE MEDIA AND ARGOUML VARIANTS

| Case study | Release | Release description |
|---|---|---|
| Mobile media | 1 | The first release of mobile photo software implements the core system "*i.e.,* mobile photo core". |
| | 2 | The second release of mobile photo software implements the exception handling "*i.e.,* exception handling included". |
| ArgoUML | 1 | The first release of ArgoUML software supports all standard UML diagrams except sequence diagram "*i.e.,* only sequence diagram disabled". |
| | 2 | The second release of ArgoUML software supports all standard UML diagrams except use case diagram "*i.e.,* only use-case diagram disabled". |

The different case studies show different sizes: ArgoUML (large product variants), mobile media (medium product variants), and drawing shapes (small product variants). However, the different complexity levels display the scalability of Juana to dealing with such product variants. ArgoUML and mobile media software variants are presented in Table III characterized by metrics LOC (Lines of Code), NoP (Number of Packages), and NoC (Number of Classes).

TABLE III
ARGOUML AND MOBILE MEDIA SOFTWARE PRODUCT VARIANTS

| Product variants | Product Description | LoC | NoP | NoC |
|---|---|---|---|---|
| ArgoUML | Only sequence diagram disabled | 114,969 | 86 | 1,608 |
| | Only use-case diagram disabled | 117,636 | 87 | 1,625 |
| Mobile media | Mobile photo core | 936 | 10 | 16 |
| | Exception handling included | 1,213 | 15 | 25 |

The AOC-posets in Figure 5 shows the evolution scenarios in mobile media at the package and class levels.

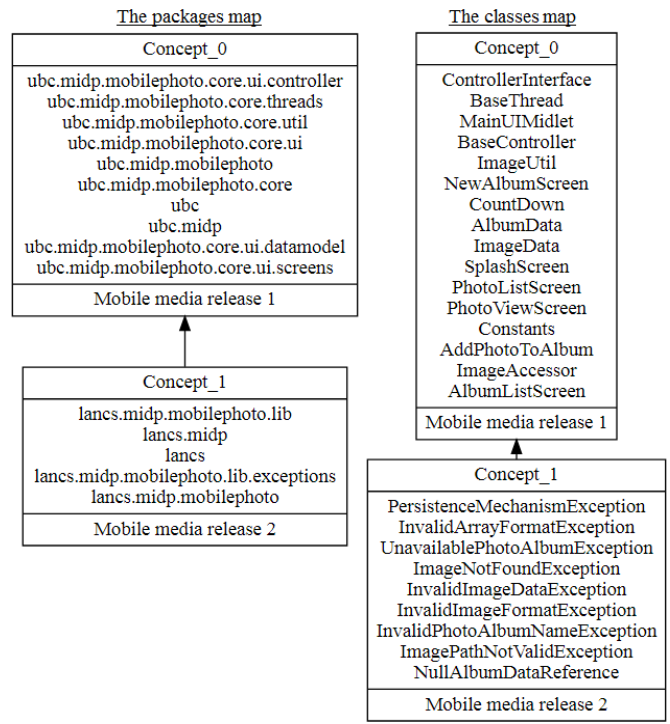

Fig. 5. The packages and classes map for mobile media software variants

Based on the mobile media documents, Juana detects the evolution scenario at the source code level in two versions of mobile media in an accurate manner. In Figure 5, the intent of the most general concept (*i.e.,* Concept_0) holds package and class names that are common to all products. The intent of the remaining concept (*i.e.,* Concept_1) holds a set of package and class names unique to one product. The extent of Concept_1 is the product name holding these identifier names in its source code.

Algorithms for building AOC-posets are presented in [24, 25] and all AOC-posets in this paper built using eRCA[5] tool [12, 26]. All mobile media maps are available on Juana webpage [21]. Juana performed an evaluation of the execution time (in

---
[5] http://code.google.com/p/erca/

*milliseconds*) of its algorithms using the mobile media and ArgoUML software. Table IV presents the execution time for each case study. In Juana's approach, the identifier name is mentioned once in the map and there is no repetition because the goal is to discover the differences in the code level between the two programs. Juana's prototype, static code parser, and all code maps are available on Juana webpage [21].

TABLE IV
EXECUTION TIME OF JUANA APPROACH ACROSS VARIOUS CASE STUDIES

| Case study | Map type | Execution time (in ms) |
|---|---|---|
| Mobile media | All identifiers map | 1398 |
| | Packages map | 1144 |
| | Classes map | 1257 |
| | Methods map | 1368 |
| | Attributes map | 1330 |
| ArgoUML | All identifiers map | 895092 |
| | Packages map | 5033 |
| | Classes map | 18569 |
| | Methods map | 338196 |
| | Attributes map | 29566 |

The AOC-poset in Figure 6 shows the evolution scenario in the ArgoUML at the package level. The top concept of the AOC-poset (*i.e.*, Concept_2) presented in a simplified form (*i.e.*, too large). Based on the ArgoUML documents, Juana identifies the common and unique software identifier names in two versions of ArgoUML software in a precise manner. All ArgoUML maps are available on Juana webpage [21]. The selected case studies are used to assess many studies in the field of software engineering. Also, the selected case studies are well documented, and their evolution scenarios are available for comparison to Juana's results and validation of the approach.

Results show that Juana's approach is able to identify common and unique identifier names across two software product variants. The software identifiers map is very useful to detect the evolution scenarios at the source code level. The generated maps can be used to improve existing feature location techniques [27, 28, 31].

Results have found that Juana's map showed different evolution scenarios between two releases. First, the added scenario, in this case, the software identifier name did not exist in the initial version but exists in the current version. Second, the removed scenario, where the software identifier name existed in the initial version but does not exist in the current version. In the case of an unchanged scenario, the software identifier name exists in both releases and did not change [29]. Software identifiers are important resources to analyze software systems [30]. Thus, a software identifiers map is extracted from two versions of a software system. In addition, the software identifiers map is important for software developers to understand the evolution scenarios for legacy systems when the software documents are missing. For example, based on the identifiers map, some identifier names that existed in the first release are deleted from the second version, and new identifier names are added to the second version to fix bugs (*e.g.*, mobile media) or to add some functionalities (*e.g.*, ArgoUML).

To evaluate the suggested approach, the author performs a simple evaluation with ten Java developers as participants. Upon starting the evaluation, each participant was asked to see the identifiers map of ArgoUML and mobile media. Then, each participant was asked if he/she was felt such graphs will be helpful for them to understand what happens between two releases. All participants were felt that the extracted map was very important as the changes between the two versions were very precise.

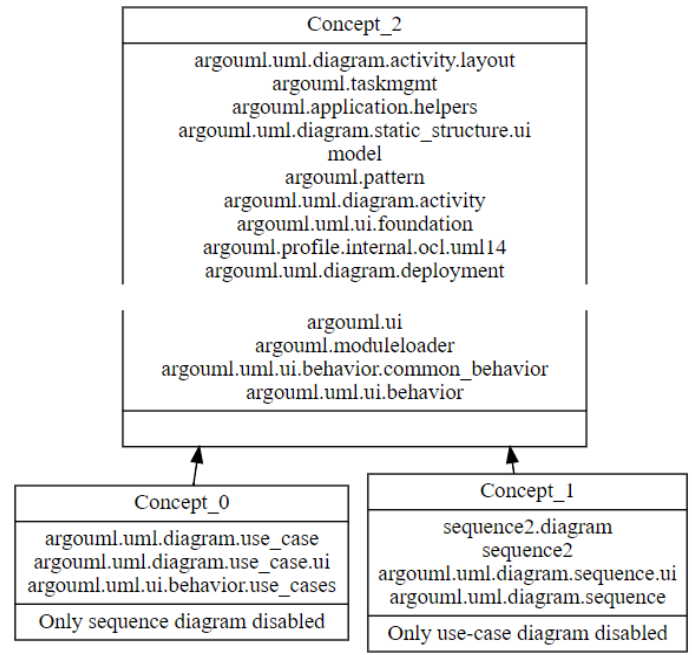

Fig. 6. The packages map for ArgoUML software variants

Juana's approach has been evaluated by three metrics: precision, recall, and F-Measure [19]. All metrics have values between 0 and 1. Table V presented the evaluation metrics of Juana's approach.

TABLE V
EVALUATION METRICS: PRECISION, RECALL, AND F-MEASURE

| Precision = \|{relevant IN} ∩ {retrieved IN}\| / \|{retrieved IN}\| |
|---|
| Recall = \|{relevant IN} ∩ {retrieved IN}\| / \|{relevant IN}\| |
| F−Measure = 2 × [(Precision × Recall) / (Precision + Recall)] |
| IN stands for identifier names |

Results have shown that precision, recall, and F-Measure value is one of all mined identifier maps thanks to our approach that identifies common and unique identifier names by using FCA. Thus, all identifier names of the retrieved map are relevant, and all relevant identifier names are retrieved. Table VI illustrates the obtained results of some identifier names from case studies (*i.e.*, package names). Since the extracted map contains the same identifier names as in the original code, the approach is accurate and only retrieves the identifier names as they are in the software code.

Results have displayed that all evaluation metrics appear high for the extracted identifiers map. This means that all extracted identifier names on the map are correct and relevant. As concepts of the AOC-posets are well-organized, the intent of the top concept holds identifier names that are common to all software variants. The intents of the two remaining concepts hold sets of identifier names unique to one variant and correspond to the implementation of one or more functionalities. The extent of each of these concepts is the product variant name containing these identifier names in its source code (*cf.* Figure 6).

TABLE VI
PACKAGE NAMES MINED FROM CASE STUDIES

| Case study | * | ** | Evaluation metrics | | |
|---|---|---|---|---|---|
| | | | Precision | Recall | F-Me. |
| ArgoUML | 90 | 90 | 1 | 1 | 1 |
| Mobile media | 15 | 15 | 1 | 1 | 1 |
| * | The number of package names in product variants code | | | | |
| ** | The number of the package names on the map | | | | |
| Statistical information | | | | | |
| ArgoUML | | | | | |
| The common package names | | | | | 83 |
| The unique package names for "only use-case diagram disabled" | | | | | 4 |
| The unique package names for "Only sequence diagram disabled" | | | | | 3 |
| Total number of package names | | | | | 90 |
| Mobile media | | | | | |
| The common package names | | | | | 10 |
| The unique package names for "Mobile photo core" | | | | | 0 |
| The unique package names for "Exception handling included" | | | | | 5 |
| Total number of package names | | | | | 15 |

The AOC-poset in Figure 7 displays the evolution scenario in the ArgoUML at the class level. Also, the top concept of the AOC-poset (*i.e.,* Concept_2) offered in a simplified form (*i.e.,* too large). The extracted identifiers map precisely shows the differences at the code level among software product variants. Results have shown that the identifiers map displays all the names of the identifiers that are in the original code of the software products. Thus, Juana helps software engineers understand the evolution scenarios across software systems.

The *threat to the validity* of Juana is that software engineers might not use the same vocabularies to name software identifiers across software variants. As an example, product A contains "salary" and "income" classes, while product B contains "employeeSalary" and "tax" classes. In this case, "salary" and "employeeSalary" are different names for the same software class. Thus, Juana might not be reliable (or should be improved with other techniques) in all cases to detect evolution scenarios across product variants. Also, Juana considers only the Java software systems. Thus, the prototype works only with Java software systems.

## VI. CONCLUSION AND FUTURE WORK

This paper focused on detecting common and unique software identifier names of software product variants realized via code cloning. Juana's approach aimed to find those identifier names that are present in one variant of the software and absent in another. The software family is usually well documented but detecting the common and unique identifier names in a given software variant still a challenging task and imprecise in many cases. In this paper, Juana's approach was based on FCA to identify the common and unique identifiers from the OO source code of two software product variants. In fact, developers can use this approach to understand the changes that have occurred during program evolution. The novelty of Juana is the exploiting of software identifier names to understand the software evolution scenarios across the product family. The proposed approach was applied to three case studies, and the results proved the validity and accuracy in identifying the changes that occurred during program evolution by comparing the result of Juana with available documents for each case study. For future work, Juana's approach will be extended by comparing more than two software variants to identify common and unique software identifier names. Also, Juana's approach plans to apply the tag cloud visualization technique [32 – 34, 47] on common and unique identifier name blocks to present the most frequent words in those blocks to software engineers.

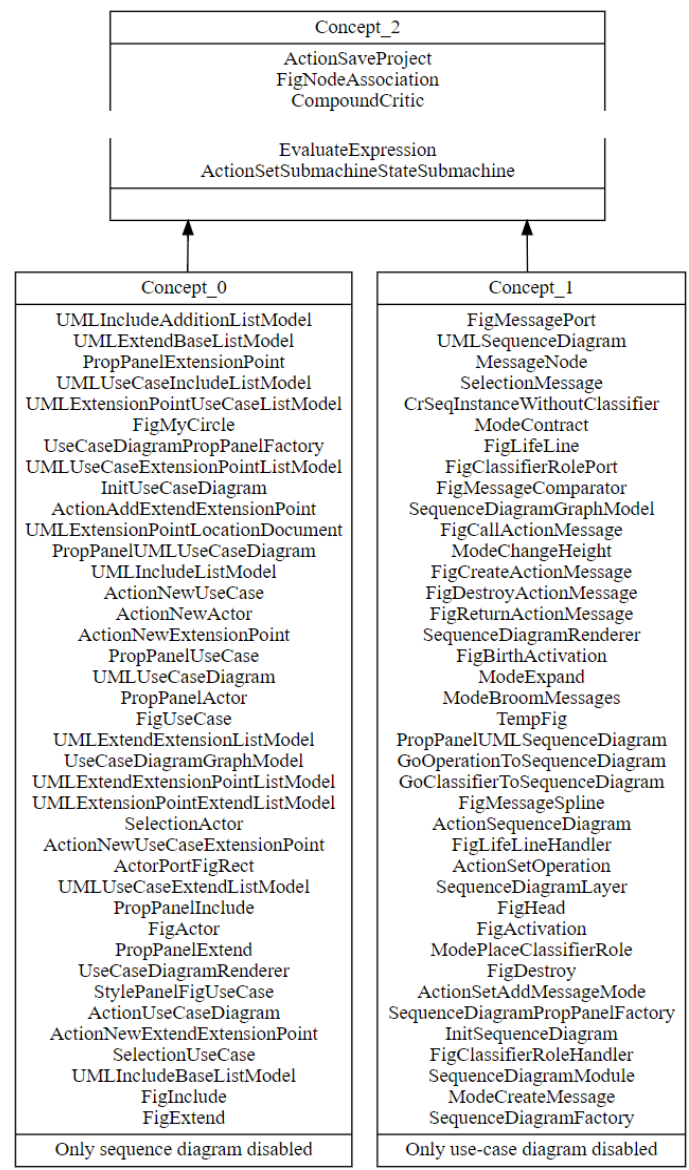

Fig. 7. The class names map for ArgoUML software variants

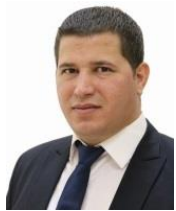
**Ra'Fat Al-Msie'Deen** is an Associate Professor at Mutah University since 2014. He received his PhD in Software Engineering from the University of Montpellier 2, Montpellier – France, in 2014. He received his MSc in Information Technology from the University Utara Malaysia, Kedah – Malaysia, in 2009. He got his BSc in Computer Science from Al-Hussein Bin Talal University, Ma'an – Jordan, in 2007. His research interests include software engineering, software product line engineering, and formal concept analysis.

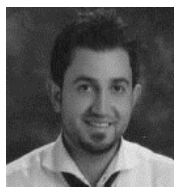
**Anas Blasi** is an Associate professor in the CIS department at Mutah University. He earned the MSc in Computer Science from University of Sunderland (England) in 2010, and the Ph.D. in Computer and systems Science from the State University of New York at Binghamton (USA) in 2013. Dr. Blasi research area is focusing on AI, Data Mining, Data Science, Machine Learning, Optimization algorithms, Fuzzy logic, and EDM. He has published several papers in reputed journals and conferences.